\documentstyle[prb,twocolumn,aps]{revtex}

\begin{document}

\twocolumn[\hsize\textwidth\columnwidth\hsize\csname
@twocolumnfalse\endcsname \title{NbSe$_3$: Effect of Uniaxial Stress on the
Threshold Field and Fermiology}\author{Jahyong Kuh, Y.T. Tseng, Keith
Wagner, Jay Brooks, G.X. Tessema, and M.J.Skove}\address{Department of
Physics, Clemson University, Clemson, SC,
29634-1911}\date{\today}\maketitle\begin{abstract}We have measured the
effect of elastic strain $\epsilon $ on the threshold field $E_T$ for the
motion of the higher temperature charge density wave(CDW) in NbSe$_3$. We
find that $E_T$ exhibits a critical behavior, $E_T\sim(1-\epsilon /\epsilon
_c)^\gamma $ where $\epsilon _c$ is about 2.6\%, $\gamma $ $\sim $1.2. This
expression remains valid over more than two decades of $E_T$, up to the
highest fields of about 1.5 kV/m measured using pulse techniques. Neither
$\gamma $ nor $\epsilon _c$ is very sensitive to the impurity content of the
sample. The transition temperature is linear with $\epsilon $, and
$dT_P/d\epsilon $ = 10 K/\% shows no anomaly near $\epsilon _c$. The slope of
the narrow band noise frequency vs. the CDW current does not change
appreciably with $\epsilon $. Shubnikov-de Haas measurements show that the
extremal area of the Fermi surface decreases with increasing $\epsilon $. We
conclude that there is a very intimate relationship between pinning and the
Fermiology in NbSe$_3$.\end{abstract}\pacs{Valid PACS appear
here.{\tt$\backslash$\string pacs\{\}} should always be input,even if
empty.}]

{\bf Introduction:}

The recent discovery of the Aharonov-Bohm effect exhibited by the sliding
CDW in NbSe$_3$ has revived interest for the field of CDW \cite{1} .
Non-linear conductivity is the outstanding characteristic of charge density
wave materials \cite{2} . The presence of a threshold field $E_T$, above
which the resistance decreases, is the signature that the CDW can be made to
move under a small electric field \cite{3,4}. The dependence of $E_T$ on
temperature (T), number of impurities ($n_i$) \cite{5,6}, contact position,
size \cite{7}, pressure\cite{8} and uniaxial stress \cite{9,10}, has been
extensively reported. Here we report further studies on the effect of
elastic, uniaxial stress $\sigma $ on $E_T$ for the upper CDW in NbSe$_3$.
This paper will show that $E_T\sim 1/$ 
\mbox{$\vert$}%
$\sigma -\sigma _C$ 
\mbox{$\vert$}%
, where $\sigma _C$ $\approx $ 260 GPa and that this is related to the
change in Fermiology as shown from low temperature Subnikov-de Haas (SdH)
studies.

\section{Experimental techniques and results}

\subsection{Samples}

The experiments were conducted on nominally pure as well as Fe doped NbSe$_3$
samples. Fe doping was achieved by mixing either 4.7\% or 7\% of Fe in the
starting materials; about a tenth that much doping is expected in the
resulting whiskers\cite{11}. Samples of medium purity with a RRR of 70 were
grown in house; the high purity samples, RRR 
\mbox{$>$}%
200, were provided by R.E. Thorne. The samples were mounted on a stressing
device described elsewhere \cite{12}. Uniaxial stress was applied along the
needle axis, the {\bf b} crystal axis. The strain $\varepsilon $ was
directly measured, and can be converted to stress using the Young's modulus
, S$_{22}\approx $ 100 GPa\cite{10}. Four electrical contacts were made
using conducting silver paint. Epoxy overlaid these contacts and formed the
mechanical grips. Typical sample dimensions were 2,000x 20 x1 $\mu $m$^3$.

\subsection{Effect of $\varepsilon $ on the Upper CDW}

At low strain and low fields, the threshold field was determined using the
conventional lockin or $dV/dI$ technique. At high $\varepsilon $, where high
electric fields are required to reach $E_T$, the pulse method was used. The
duty cycle was less than 1\%; with typical pulse width of 10 $\mu $s, and
period 1 ms. The pulsed current and voltage were measured using a two
channel boxcar signal averager, EG\&G model 162. In this case, $E_T$ was
estimated from the plot of the chordal resistance R vs $E$ or from the
numerical derivative $\Delta $V/$\Delta $I. Previous studies \cite{9,10}
have shown that uniaxial stress affects $E_T$ indirectly by $\varepsilon $
induced changes in $T_P$ and directly by enhancing the pinning strength. It
was shown that the indirect effect can be disentangled by conducting the
experiments at a constant reduced temperature $t=T/T_p(\varepsilon )$ where $%
T_p(\varepsilon )$ is defined at the peak in $dR(\varepsilon )/dT$. Constant 
$t$ = 0.70 was achieved by adjusting $T$ for each value of $\varepsilon $ .
This value of $t$ corresponds to the minimum value $E_{min}$ on the $E_T$
vs. $t$ curve. It was previously shown that $E_{min}$ is proportional to the
impurity concentration\cite{6,7} and assumed to correspond to bulk pinning
rather than contact pinning. Although this paper is devoted to the study of
the effect of $\varepsilon $ on $E_{min}$, for the sake of simplicity we
will refer to it as the threshold field, or simply $E_T$.

\begin{figure}[tbp]
\caption{ E$_T$ vs $\varepsilon $ for a nominally pure sample. The inset
shows a semilog plot of the same data.}
\label{1}
\end{figure}

Figure 1 shows a typical plot of $E_T$ vs. $\varepsilon $ for an arbitrarily
selected sample. $E_T$ increases weakly at low strain and diverges near $%
\varepsilon _c$ = 2.6 $\pm \ 0.3\%$. The semi-log plot of the same data
shown in the inset indicates that $E_T$ increases faster than a single
exponential. In the few cases where we were able to pull beyond 2.6\%
strain, $E_T$ exhibited a peak, decreasing above 2.6 $\pm $ 0.3\%. Figure 2
shows the strain dependence of $T_P$. $T_P$ decreases linearly with
increasing $\varepsilon $ up to 3 \% at a rate $dT_P/d\varepsilon $ = 10
K/\% . There is no apparent feature around $\varepsilon $ = 2.6 \%, where $%
E_T$ diverges. The inset in Figure 2 shows a plot of $R$ vs $T$ for
different values of $\varepsilon $. Note that the resistance anomaly $\Delta
R(\varepsilon )$ = $R_p(\varepsilon )-R_{fit}(\varepsilon )$ is independent
of $\varepsilon $, where $R_p(\varepsilon )$ corresponds to the peak
resistance for a given $R(\varepsilon )$ vs $T$ plot, and $R(\varepsilon
)_{fit}$is the linearly extrapolated resistance at $T$ peak from above 150
K. This result suggests that the CDW conductance $(G_{cdw}\sim n_{cdw}e\mu $
where $e$ is the charge of the electron and $\mu $ the cdw mobility) at very
large electric fields is independent of $\varepsilon $, which in turn
implies that the fraction of condensed electrons $n_{cdw}$ does not change
appreciably with $\varepsilon $. This is consistent with narrow band noise
measurements which showed almost no change in the slope of the CDW current
vs the narrow band frequency $(dI_{cdw}/dF)$ with $\varepsilon $\cite{13}.

\begin{figure}[tbp]
\caption{ The CDW transition temperature T$_{p1}$ vs. strain. T$_{p1}$
decreases linearly with $\varepsilon $ up to $\varepsilon $ = 3 \%. Typical
R($\varepsilon $) vs. T plots are shown in the inset.}
\label{2}
\end{figure}

\subsection{Effect of $\varepsilon $ on the Fermi Surface}

This section looked for a connection between the effect of $\varepsilon $ on
the Fermiology and the results reported in the previous section. The effect
of $\varepsilon $ on the dominant frequency of the Shubnikov-de Haas
oscillations is reported. The magnetic field B$_a$ was parallel to the ({\bf %
b,c}) plane and perpendicular to the smallest extremal area of the Fermi
surface, with typical frequency of 0.28 MG at $\varepsilon =0$ \cite{13,14}%
{\bf .} Two methods were used. The first method is the conventional method, $%
B_a$ was increased slowly with the sample under constant strain. In the
second method, $B_a$ was constant while sweeping $\varepsilon $. The
experiments were conducted at constant $T$ between 3.0 and 4.2 K.

Figure 3a shows a typical plot of R vs H obtained using the conventional
method, the inset shows $dR/d(1/H)$ vs $1/H$ . The extremal area $A$ was
estimated from a plot of $n$ vs $1/H$ for each value of $\varepsilon $.
Figure 3b shows that $A$ decreases nearly linearly with increasing $%
\varepsilon $. A detailed study of the effect of uniaxial stress on the
Fermi surface will be reported elsewhere. In this paper we note that
uniaxial stress suppresses $A$ linearly at the rate of 0.09 MG/\%,
suggesting that the whole pocket would be wiped out for $\varepsilon \leq $
3\%. A study of the strain dependence of the conductance at low temperature
shows that 90\% of the conductance is wiped out for $\varepsilon \approx $
2.6\%. This suggests that this pocket plays a predominant role in the normal
state conductance of NbSe$_3$ at low $T$.

\begin{figure}[tbp]
\caption{ Fig 3a shows a typical plot of R vs H which exhibits a the
Shubnikov-de-Haas oscillations. The derivative dR/d(1/H) vs 1/H is shown in
the inset. In fig3b the extremal area, in units of kG, decreases smoothly
with increasing $\varepsilon .$}
\label{3}
\end{figure}

The second method is equivalent to fixing the Landau tubes and shrinking the
Fermi surface through them under the influence of $\varepsilon $. This leads
to oscillations in the $R$ vs $\varepsilon $ plots as shown in Fig. 4a. A
systematic study of $R$ vs $\varepsilon $ for different values of $B_a$
allows us to follow the strain and the field at which a given Landau{\bf \ }%
tube is crossed. The results are summarized in Fig. 4b., which shows a plot
of $\varepsilon $ vs $B_a$ for each Landau tube identified by the integer
next to its curve. The trajectory of a given Landau tube is nearly linear.
This is consistent with the linear relationship between $A$ and $\varepsilon 
$ observed using the conventional technique. The solid lines in the figure
are a guide to the eye. Note that at $B_a$ = 0 T, all the lines converge to
nearly the same $\varepsilon _c^H$ $\simeq $ 2.6 \%. This suggests that this
piece of the Fermi surface would be wiped out at about 2.6 \%. Below we will
also show that $\varepsilon _c^H$ is equal to the critical strain $%
\varepsilon _c^{E_T}$ derived from the critical plot of $E_t$ of the upper
CDW. 
\begin{figure}[tbp]
\caption{ Fig 4a shows the oscillatory R vs $\varepsilon $ plots for B= 5.4\
Tesla. The oscillations are attributed to the intersection of the Landau
tubes with the shrinking Fermi surface. Figure 4b is a representation of
this intersection in the ($\varepsilon $, B) space. The integers in the box
correspond to the indices of the Landau levels.}
\end{figure}

\section{\bf Discussion}

Possible pinning mechanisms of the CDW are: bulk impurity pinning as
discussed by Fukuyama-Lee-Rice\cite{16} (either strong or weak),
commensurability pinning by the underlying lattice, or pinning by other
defects such as surfaces, dislocations or contacts. The results in Fig. 3
could be due to one or a complex combination of the following effects: (1)
strain induced enhancement of the weak impurity pinning potential; (2)
strain induced crossover from weak pinning to strong pinning; (3) strain
induced incommensurate to commensurate transition; or (4) strain induced
enhancement of contact pinning. We now discuss each one of these effects
separately, in inverse order of their likelihood.

Stress induced enhancement of contact pinning is very unlikely. If this were
the case, one would also expect a similar stress induced enhancement of $E_T$
for the lower CDW. However, previous studies have shown that stress does not
enhance $E_T$ for the lower CDW\cite{9,10}. In addition, Y. Tseng et al.
have shown \cite{9} in the case of the upper CDW, $E_T$ can be separated
into two components, one attributed to contact pinning, and the other to
bulk impurity pinning. They have argued that uniaxial stress does not
enhance contact pinning. It also seems unlikely that uniaxial stress can
affect surface pinning to that extent; if it did, our thinner samples would
have shown a stronger effect.

The FLR model considers two possible kinds of impurity pinning: strong
pinning and weak pinning. Several experiments indicate that pinning in NbSe$%
_3$ is due to weak pinning\cite{6,7}. Stress-induced crossover from weak
pinning to strong pinning could be considered, in which case one would
expect to see a change in the exponent $\gamma $ with $\varepsilon $. $%
\gamma $ is defined in the next paragraph. The results show that $\gamma $
is independent of $\varepsilon $ and rule out this possibility as well. This
is also supported by the fact that the same $\gamma $ is obtained for the
samples doped with Fe, which may be considered as a strong pinning impurity.

\begin{figure}[tbp]
\caption{ A critical plot of E$_T$ for five different samples. The full
triangles and the crossed squares correspond to Fe doped samples, 0.7\% and
0.47\% respectively. The normalized threshold e$_T$ = E$_T$($\varepsilon $)/E%
$_T$(0) is shown in the inset. Note that all five set of data fall on the
same line. }
\label{5}
\end{figure}
Experimental search for an incommensurate-commensurate transition (ICT) in
CDW systems has not provided any clear evidence for these effects\cite{17}.
A stress-induced ICT would lead to changes in $dT_P/d\sigma $ as well as
soliton-like behavior near commensurability. Figure 2 shows that $T_P$ does
not show an anomaly near $\varepsilon _c$. Further, according to Fisher and
Fisher\cite{18} the approach to commensurability should behave critically
with an exponent of $\frac 12$ (for 2-D) or be logarithmic. Figure 5 shows
such a critical plot of $E_T$ vs $(1-\varepsilon /\varepsilon _c)$ in a
log-log scale. A plot for the normalized threshold field{\bf \ 
\begin{equation}
e_T=E_T/E_{T_0}=(1-\varepsilon /\varepsilon _c)^\gamma  \label{et}
\end{equation}
}where $E_{T_0}$ is the threshold field at zero strain , and $\varepsilon _c$%
and $\gamma $ are adjustable parameters is shown in the inset. In the
following we will replace $\varepsilon _cwith$ $\varepsilon _c^{E_T}$ in
order to differentiate it with the critical strain defined from the
fermiology study. Note that the results for five different samples with
different impurity content, and Fe impurity fall along the same line, with
nearly the same parameters. A list of the values of $\varepsilon _c^{E_T}$
for different samples is shown in Table 1. Although the figure is in
qualitative agreement with Fisher and Fishers prediction that $E_T$ should
behave critically in an ICT, the exponents are not in quantitative agreement
with the model. Our larger samples are most likely 3-D, and the exponent is
not 1/2; therefore the critical behavior cannot be explained by a simple
approach to commensurability. On the other hand, an argument in favor of ICT
can be made based on the divergence of $E_T$. Since commensurability pinning
is much stronger than impurity pinning, $E_T$ is much more sensitive to
stress-induced ICT than $T_P$ is. This issue could be resolved using
structural studies as a function of strain.

{\bf Table 1:} The fitting parameters $\varepsilon _c^{E_T}$ and $\gamma $
are shown together with other relevant parameters such as the nominal
purity, the RRR, and the threshold field E$_T$. Only the nominal Fe doping
levels are given.

\begin{tabular}{|l|l|l|l|l|}
\hline
Sample & RRR & $\varepsilon _c^{E_T}\%$ & $\gamma $ & E$_T$ (V/m) \\ \hline
Pure & 250 & 2.6 & 1.58 & 8.0 \\ \hline
Pure & 200 & 3 & 1.23 & 11 \\ \hline
Pure & - & 2.6 & 1.23 & 6.3 \\ \hline
Pure & - & 2.6 & 1.66 & 10 \\ \hline
4.7\% Fe & - & 3.2 & 1.14 & 28 \\ \hline
4.7 \% Fe & - & 2.6 & 1.08 & 31 \\ \hline
7\% Fe & - & 2.7 & 1.23 & 63 \\ \hline
\end{tabular}

According to the FLR theory\cite{16}, the threshold field for weak pinning
is given by: 
\begin{equation}
eE_T\lambda \propto \frac{\Delta ^2}{E_F}(\xi _x,\xi _y,\xi _zn_i^2)V_0^{(%
\frac 4{4-D})}  \label{eEt}
\end{equation}
where $\lambda $ is the wavelength of the CDW, $\xi _x,\xi _y,$and $\xi _z$
are coherence lengths for the CDW amplitude, $E_T$ the threshold field, e
the electric charge, $E_F$ the Fermi energy, $V_0$ the impurity potential
and $D$ the dimensionality\cite{16}. Stress could effect any or all of the
parameters in Eq. 2. However, for the sake of simplicity we will discuss
separately the terms that are susceptible to change with $T_P$. In the
conventional BCS model the CDW gap is proportional to $T_P$, $\Delta $/$T_P$
= 4.8 for NbSe$_3$\cite{1}. As in previous pressure work\cite{8}, uniaxial
stress induced enhacement of the electron-phonon coupling constant could be
considered. However, it would take more than an order of magnitude of change
in order to account for our results. On the other hand if, in a first
approximation, one assumes that this ratio is not affected by $\varepsilon $%
, $\Delta $ would decrease with $T_P$ which would lead to a decrease in $E_T$%
, contrary to our results. Another{\bf \ }possibility is a strain-induced
decrease in $E_F$. But the normal state conductivity around $T_P$ is a weak
function of $\varepsilon $, even up to 3\%, suggesting that $E_F$ is not
strongly affected by $\varepsilon $. One likely possibility is that $V_0$ is
strongly affected by stress. Suppose there is a stress-induced tuning of the
matching between the phase and wavelength of the CDW and the Friedel
oscillations \cite{19}. Then, although the changes in $E_F$ due to the
vanishing of this small pocket could be negligible, it could be sufficient
to lead to a rapid increase of $V_0$. This mechanism would be independent of
the type of impurity and concentration, in agreement with our experiments.
One other possibility is that the pocket screens the impurity, and V$_o$
increases when the pockets disappears. Below we will discuss the difference
between the upper and lower CDW.

In a study of the combined effect of magnetic field and strain, Parilla et
al. \cite{20} have shown that uniaxial stress and $\mu _0H$ act on the same
piece of the Fermi surface. This was confirmed by Y.T. Tseng et al. \cite{21}
who observed pronounced effect of strain on the resistance and thermopower
of NbSe$_3$ below 59 K. Jianhui et al.\cite{22} have conducted NMR
experiments to study the density of states on the different chains in NbSe$%
_3.$ Although magnetic fields effects on the ohmic regime, below $E_T$ , are
much more pronounced below the second transition than below $T_{p1}$, their
results show that most of the changes in FS is due to changes in density of
states on the chain associated with the upper CDW rather than the lower CDW.
This supports the notion that the strain-induced changes in $E_T$ of the
upper CDW are associated with changes in the Fermi surface most closely
associated with the chain corresponding to the upper CDW. The relatively
small effects on the density of states associated with the lower CDW could
account for the rather weak effect on the $E_T$ of the lower CDW. \medskip

\section{\bf Conclusion}

\medskip It was previously reported that uniaxial stress enhances $E_T$ for
the upper CDW in NbSe$_3$. In this paper we report a systematic study of the
effect of $\varepsilon $ on $E_T$, $T_P$, and the Fermi surface of this
compound. We show that the divergence of $E_T$ near 2.6\% strain is
intimately related to stress induced changes in fermiology. We propose that
the two most likely possibilities for this phenomena are (1) a stress
induced incommensurate to commensurate transition (2) or more likely an $%
\varepsilon $ driven matching of the Friedel oscillations and the CDW
oscillations of the upper CDW. Structural studies under stress should give a
further insight into the subject.

\strut

Acknowledgements: The authors would like to thank Prof. John C. MaCarten for
his valuable comments and suggestions, and R.\ E. Thorne for his high purity
samples. This work was supported by the National Science Foundation:
DMR\#9312530.

{\bf I}

\end{document}